\documentclass[aps,prl,twocolumn,floats,showpacs]{revtex4}
\usepackage{graphicx}
\usepackage{color}
\usepackage{bm}
\usepackage{ifthen}
\usepackage{ifpdf}

\usepackage{latexsym}
\usepackage{amsmath}
\usepackage{amssymb}
\usepackage{bm}
\usepackage{wasysym}

\ifpdf
\usepackage{graphicx}
\usepackage{epstopdf}
\else
\usepackage{graphicx}
\usepackage{epsfig}
\fi
\newcommand{\hide}[1]{}

\begin{document}
\title{Metal-Quantum Dot-Topological Superconductor Junction: \\
Kondo correlations and Majorana Bound States }
\author{ A. Golub,  I. Kuzmenko, and Y. Avishai}
\affiliation{ Department of Physics, Ben-Gurion University, Beer Sheva 84105 Israel\\   }
\begin{abstract}
Electron transport through [normal metal]-[quantum dot]-[topological superconductor] junction is studied and reveals interlacing physics of Kondo correlations with two Majorana fermions bound state residing on the opposite edges of the topological superconductor. 
When the strength of the Majorana fermion coupling exceeds the temperature $T$, this combination of Kono-Majorana fermion physics can be observed: The usual peak of the temperature dependent zero biased conductance $\sigma(V=0,T)$ splits and the conductance has a {\it dip} at $T=0$. The height of the conductance side-peaks decreases with magnetic field.
\end{abstract}
\maketitle
\noindent
\underline {{\it Introduction}:} Recent theoretical investigation of topological materials reveals that topological superconductors can host Majorana fermions\cite{fu,sato,oreg,das,alicea,potter,qi,been}. Specifically, Majorana bound states (MBS) reside  at the ends of a 1-D topological superconductor (TS).
Interest in the physics of Majorana quasiparticles is due to their non-Abelian statistics\cite{nayak}.  Hopefully,
 MBS can be realized on the 1-dimensional edge of a 2D quantum spin Hall insulator with proximity induced superconducting gap \cite{kitaev,fu2}, or at the ends of a one-dimensional semiconducting wire with spin-orbit coupling, in proximity with an s-wave superconductor \cite{oreg, das}. In both cases a 1D  TS is formed. Two Majorana fermions can form a neutral Dirac fermion, but detecting it requires non-local measurements. Yet, it was suggested \cite{demler,law,linder} that  MBS can be probed in tunneling process, namely, a local measurement that is sensitive to interference between various MBS. \\
 Motivated by the above analysis, we consider electron tunneling through an N-QD-TS junction composed of normal metal lead, quantum dot and 1D topological superconductor. As was already noted, the TS hosts two MBS on its ends \cite{alicea2,flens}. Consequently, within a reasonable approximation,
 the tunneling problem is reduced
 to that of transport  in an N-QD-MBS junction. Our interest is focused on the
 interrelation between the Kondo
 physics prevailing in the quantum dot and the MBS physics prevailing on the TS.
 The analysis is naturally divided into the weak ($T \gg T_K$) and strong  $(T<T_K)$ coupling regimes, where $T_K$ is the Kondo temperature.
 The effect of a magnetic field acting on the dot (through a Zeeman term)
 is also analyzed since it is an important for obtaining the 1D TS. We start our analyzes with the weak coupling limit which can be studied by the perturbation theory\\
\underline{{\it The Hamiltonian}} of the junction
\begin{equation}\label{H}
    H=H_0+H_M+H_K+BS_z,
\end{equation}
 includes the following components: 1) $H_0$ for the normal metal lead,  held at bias voltage $V$. 2) The Kondo part $H_K=H_{LK}+H_{KM}$
 expressed in terms of dot-lead and dot-Majorana fermion exchange interactions $H_{LK}$ and $H_{KM}$,
obtained by applying  the  Schrieffer-Wolff  transformation on the
$U/t \to \infty$ Anderson Hamiltonian with hopping energy $t$, level energy $\varepsilon$ and width $\Gamma=2 \pi^2|t|^2 N(0) \ll |\varepsilon|$.   3) the Majorana term $H_M$ describes coupling of strength $\nu$ between two Majorana fermions
$1$ on the left end and $2$ on the right end of the TS . 4) The Zeeman energy $BS_z$ of the dot subject to an external magnetic field  $B$. \\
 Employing Nambu formalism
 in the 4 dimensional space [spin]$\otimes$ [electron-hole], we have, assuming the dot is at $x=0$,
\begin{eqnarray}
H_M&=&\frac{i}{2} \sum_{i,j=1}^2\hat {\nu}_{ij}\gamma_i \gamma_j,\label{M}\\
H_{LK}&=&\frac{J_L}{2}c^{\dagger}(0)Q\tau_zc(0), \\
H_{KM}&=&\frac{t}{|\varepsilon|}(c^{\dagger}(0)Q\tau_z\hat{V}\gamma_1+H.C).
 \label{HM}
\end{eqnarray}
 The definitions are as follows:
 $\gamma_i$ $(i=1,2)$ are Majorana fermion operators  satisfying $\gamma_i= \gamma_i^{\dagger}$, $\gamma_i^2=1$. The coupling between two MBS is given by an antisymmetric $2\otimes2$ matrix $\hat{\nu}=\nu \tiny {\left (\begin{array}{cc} 0 &1\\-1&0 \end{array} \right ) }$.
 The Kondo coupling constant to the normal metal is $J_L=2|t|^2/|\varepsilon|$,
and  $c(0)=(c_{\uparrow},c_{\downarrow},c_{\downarrow}^{\dagger},-c_{\uparrow}^{\dagger})^T$ is the normal lead electron operator at $x=0$.
In Nambu representation, the coupling  between the quantum dot and Majorana states is given by a vector $\hat{V}_i=(\lambda_{\uparrow},\lambda_{\downarrow},\lambda_{\downarrow}^{*},-\lambda_{\uparrow}^{*})^T$.
  The Pauli $\tau$-matrices act on ($c_{\sigma}$; $\lambda_{\sigma}$) and ($c_{\sigma}^{\dagger}$; $\lambda_{\sigma}^{*}$) blocks. Below we put $\lambda_{\uparrow}=\lambda,\,\lambda_{\downarrow}=-i\lambda$ \cite{flens}.
Finally, the operator $Q$ is the Numbu space extension of the exchange interaction
\begin{equation}\label{Q}
    Q=\frac{1}{4}I\otimes I+[s_xS_x+s_zS_z+s_yS_y]\otimes\tau_z,
\end{equation}
 where $\vec{s}$ is the operator of electron spin.\\
\underline{\it Keldysh actions and Green's functions (GF):} The current operator is defined as the time derivative of number operator of the normal metal lead $\hat{J}=e d\hat{N_L}/dt$ and takes a form
\begin{equation}\label{J}
   \hat{J}=\frac{ie}{\hbar}\frac{t}{|\varepsilon|}c^{\dagger}(0)Q\hat{V}\gamma_1+H.C~.
\end{equation}
Within the Keldysh  perturbation technique,
the normal metal free electron action reads,
\begin{eqnarray}\label{g}
  S_{\mathrm{lead}} &=& \frac{1}{2}\int dt\Sigma_k \bar{\hat{c}}_k g_k^{-1}\hat{c}_k,
\end{eqnarray}
where $g_k^{-1}$ is the inverse GF for lead electrons. For
later manipulations, the GF  of the lead integrated over momentum
$\bar{g} = \frac{1}{2\pi}\Sigma_k g_k $ is introduced,
$\tiny {\bar{g}=\left(
         \begin{array}{cc}
              \bar{g}^{11} & \bar{g}^<    \\
               \bar{g}^> & \bar{g}^{22}     \\
               \end{array}
                \right)}~. \label{keld}$

Here all entries are $4\times4$ diagonal matrices $\bar{g}^{11}(\omega)= \bar{g}^{22}(\omega)=\frac{i}{2}N(0) \mbox{diag}[(1-f(\omega-eV))(1,1,0,0) +(1-f(\omega+eV))(0,0,1,1)]$ and $\bar{g}^<(\omega)=-\bar{g}^>(-\omega)=
iN(0) \mbox{diag}[(f(\omega-eV)(1,1,0,0) +f(\omega+eV))(0,0,1,1)]$ where $f(\omega)$ is the Fermi distribution function. We also use the GF matrix in rotated Keldysh basis  $\tilde{g}=K\bar{g}\hat{\rho}_z K^{-1}$   (here Pauli matrices $\hat{\rho}$ act in Keldysh space and $K=\tiny{\left(
                                                                \begin{array}{cc}
                                                                  1 & -1 \\
                                                                  1 & 1 \\
                                                                \end{array}
                                                              \right)/\sqrt{2}}$).

To facilitate computation we consider the energy gap $\Delta$ of the superconductor as the highest energy scale in the problem. Then for low  applied voltage $eV<\Delta$  in the weak coupling limit $T \gg T_K$ only zero-energy Majorana operator projection of the total quasiparticle operator in the superconductor is important \cite{fu3,flens}.

\begin{figure} [!ht]
\centering
\includegraphics [width=0.5 \textwidth ]{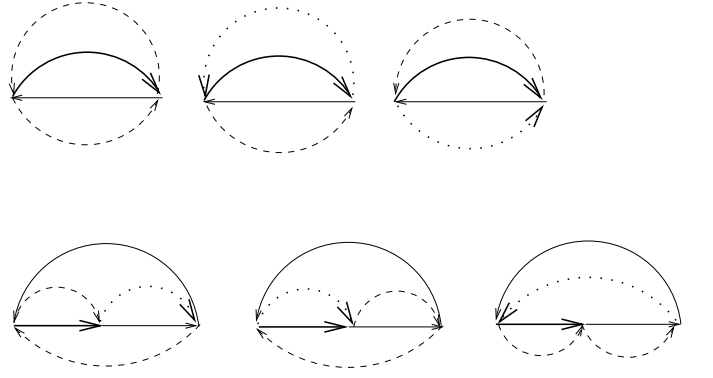}
\caption {\footnotesize{Diagrams defining the contributions to the conductance due to Kondo interaction (\ref{HM}). The upper panel represents  the second order, while lower  panel describes the third order contributions. Thick solid line - Majorana GF, thin solid lines - lead fermions GF, dashed and dotted lines - impurity spin.  Dotted line is the Majorana $F_{z}$ GF, and dashed line is the Dirac fermion $F_f$ GF. The left vertex in both sets has Keldysh index 1, other vertices have Pauli $\sigma_z$ matrix due to the doubling of the time contour in Keldysh technique.}}
\label{Fig1}
\end{figure}

To average the product of dot
spin operators we express them in terms of mixed Dirac ($f,f^{\dagger}$) and Majorana ($\eta_x,\,\eta_y,\,\,\eta_z$) fermions \cite{parcollet}:
$
  S_+ = \eta_z f^{\dagger} $; $ S_- = f \eta_z $;
  $S_z = -i\eta_x \eta_y$; $f=(\eta_x-i\eta_y)/\sqrt{2}$.
  In the diagrammatic representation of Fig.~\ref{Fig1} below we  use Keldysh GF for these fermions,
  \begin{eqnarray}
    F^{12}_f (\nu)&=& 2\pi i f(B)\delta(\nu-B),\,
    \ F^{R}_f (\nu)=\frac{1}{\nu-B+i\delta}, \nonumber\\
     F^{12}_z (\nu) &=& \pi i \delta(\nu),\,
    \ F^{R}_z (\nu)=\frac{1}{\nu+i\delta}~.
  \end{eqnarray}
The Majorana fermion action in Keldysh space follows from Eq.(\ref{M})
 to zero order in tunneling to the dot is,
\begin{eqnarray}
S_M&=&\frac{1}{2}\sum_{i,j=1,2}\int dt dt' [\gamma _i(t)]^{T}G^{-1}_{ij}(tt')\gamma_j(t').
\end{eqnarray}
Here the inverse matrix GF ,
the Keldysh components of the matrix GF for Mojorana fermions at the ends of TS
follow from Eq.(\ref{M}) are given by
\begin{eqnarray}
[G^R]^{-1}&=&\frac{1}{2}[i\partial_t-2i\hat{\nu}],\nonumber\\
ImG^{11}(\omega)&=&ImG^R(\omega)\tanh\frac{\omega}{2T},\nonumber\\
G^{<}(\omega)&=&-2f(\omega)iImG^R(\omega). \nonumber
\end{eqnarray}
\underline {{\it Non-linear conductance}:}
The calculation of the current diagrams (Fig.~\ref{Fig1} ) is straightforward. The conductance $\sigma=\frac{dJ}{dV}$ to the second order (Fig.~\ref{Fig1}) upper row) is,
\begin{eqnarray}
\sigma^{(2)}&=&\alpha W(2\nu)(1+3R(B)). \label{G2}
\end{eqnarray}
where $\alpha=\frac{\pi e^2}{2h}\frac{\Gamma|\lambda|^2}{T\varepsilon^2}$, and
\begin{eqnarray}
W(x)&=&\frac{1}{2}(\cosh^{-2}\frac{x+eV}{2T}+\cosh^{-2}\frac{x-eV}{2T}), \\
R(B)&=&\frac{1}{3}[\cosh^{-2}\frac{B}{2T}+\frac{W(B+2\nu)+W(B-2\nu)}{W(2\nu)}+\nonumber\\
&&\frac{W(B+2\nu)-W(B-2\nu)}{W(2\nu)}\tanh\frac{B}{2T}\tanh\frac{\nu}{T}]. \nonumber \\
&&
\end{eqnarray}
 In the Zero field limit $(B\rightarrow 0$) the function $R(B)\rightarrow 1$. There is a clear separation between potential and spin scattering. We also note that the temperature dependence of the non-linear conductance  is quite distinct from the standard behavior of the quantum dot between normal leads. This is due to the resonance form of the Majorana GF.
 The third order contribution to the conductance includes large logarithmic terms, that is the hallmark of the Kondo effect
  in a tunneling system with topological superconductor. Only these terms are retained,
\begin{eqnarray}
\sigma^{(3)}&=&3\alpha[\frac{\Gamma}{\pi|\varepsilon|}] W(2\nu)K(B), \label{G3}
\end{eqnarray}
where $K(B)=\kappa(B)+\kappa(-B)$ and
\begin{eqnarray}
\kappa(B)&=&\frac{1}{6}\cosh^{-2}\frac{B}{2T}L(2\nu+B)+\nonumber\\
&&\frac{W(2\nu-B)}{W(2\nu)}(L(2\nu)+L(2\nu-B))(1-\nonumber\\
&&\tanh\frac{B}{2T}\tanh\frac{\nu}{T}),
\end{eqnarray}
incorporates dominant logarithmic terms.
\begin{eqnarray}
L(x)&=&\frac{1}{2}[\ln\frac{D}{\sqrt{(x-eV)^2+T^2}}+\ln\frac{D}{\sqrt{(x+eV)^2+T^2}}]. \nonumber
\end{eqnarray}
Here $D$ denotes a high energy cut-off which corresponds to the band width in the normal lead.
The two terms ( \ref{G2},\ref{G3}) are combined to yield the total conductance in the weak coupling limit,
\begin{eqnarray}
\sigma&=&\alpha W(2\nu) \{ 1+3[R(B)+\frac{\Gamma K(B)}{2\pi|\varepsilon|} ] \} \equiv\sigma^{p}+\sigma^{ex}. \label{GG}
\end{eqnarray}
where $\sigma^{ex}$ is given by the term in the square brackets multiples by 3.
When $B\rightarrow0$,  $K(B\rightarrow0)=L(2\nu)$ and the total conductance becomes
 \begin{eqnarray}
\sigma&=&\alpha W(2\nu)[1+3(1+\frac{\Gamma}{\pi|\varepsilon|}L(2\nu))].
\end{eqnarray}
\begin{figure} [!ht]
\centering
\includegraphics [width=0.4 \textwidth ]{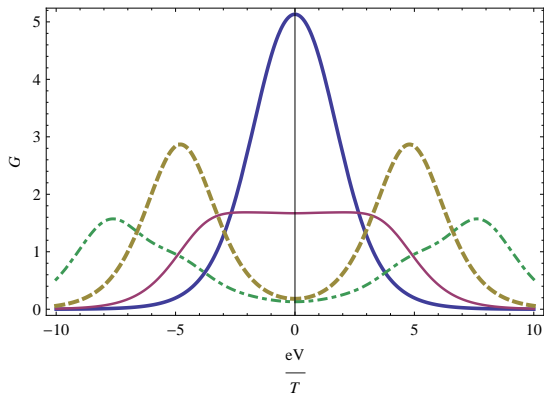}
\caption {\footnotesize {The total nonlinear conductance (\ref{GG}) $G=\sigma/\alpha$ versus applied bias. The central high peak and lower plateau correspond to  $\nu=0.4T$, $D=1000T$ but different values of magnetic field: B=0 for the peak and B=3T for the plateau. The same for the dashed and dot dashed curves, the only difference is that $\nu=2.4T$. For all four curves we assume
 $\frac{\Gamma}{2\pi\varepsilon}$=0.1.}}
 \label{Fig2}
\end{figure}
In Fig.~\ref{Fig2} the total non-linear conductance is displayed as function of the applied voltage. Two distinct peaks are resolved as the coupling energy $\nu$ of the Majorana fermions exceeds the temperature, a fact which has natural explanation. This is one of the central results of the present study since it
combines the Kondo and Majorana fermion physics, and encourages experimental
activity in searching a realization of Majorana fermions.
Under magnetic field  the heights of the peaks decrease  and reveals more complicated structure of $\sigma(V)$.
\begin{figure} [!ht]
\centering
\includegraphics [width=0.4 \textwidth ]{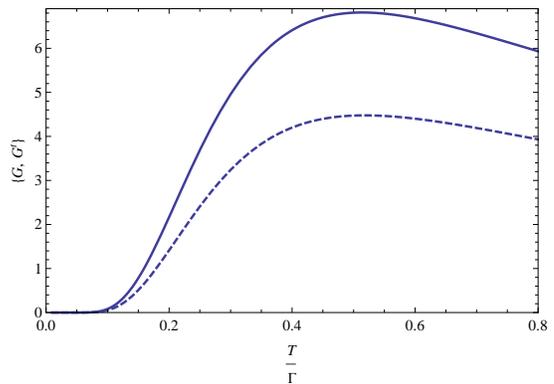}
\caption {\footnotesize{The zero biased conductance as function of temperature at zero magnetic field. $G =\sigma^{(2)}/(\alpha T/\Gamma)$ is the second order contribution (dashed line), while $G^t $ stands for the total conductance in the same units as $G$ (solid line). We take $\nu=0.4\Gamma$ and use value $\Gamma/(2\pi\varepsilon)$=0.1}.}
\label{Fig3}
\end{figure}
We can  identify Kondo correlations also by studying  the temperature dependence  of the zero bias conductance $G^t$. This is shown in Fig.\ref{Fig3}. We notice  that $G^t$ in Fig.\ref{Fig3} correlates with Fig.~\ref{Fig2}: for small temperatures where $\nu>T $  the total zero biased conductance displays a dip, which is replaced by a sharp  rise as temperature grows and exceeds $\nu$\\
\underline {{\it The strong coupling regime}:} At  $T<T_K$ we use the mean field slave boson approximation (MFSB) to estimate the zero bias tunneling  conductance. As in the weak coupling limit we neglect the continuous spectrum above the superconductor gap. In addition to the conditions $\Delta \gg T, eV,T_K$  we assume the inequality $\Gamma_S<\Gamma$ where $\Gamma_S$ is the tunneling width into the superconductor.  In this case  contribution to  conductance that originates from the continuous spectrum  is proportional (at resonance) to
 \begin{equation}\label{con}
    \sigma_c\sim 4 \left (\frac{\Gamma_S}{\Gamma} \right )^2,
\end{equation}
that is very small indeed \cite{raimondi}. The Kondo temperature $T_K=D\exp[-\pi|\varepsilon|/\Gamma]$  is determined solely by interaction with the normal metal lead. Indeed, in the weak coupling limit ($T \gg T_K$) MBS effectively represent the topological superconductor. The poor-man scaling renormalization group equations in this case read,
\begin{eqnarray}
  \frac{d J_{L}}{d\ln D} &=& -N(0)J_{L}^2, \nonumber\\
  \frac{d J_{LR}}{d\ln D} &=& -N(0) J_{L}J_{LR}, \label{RG}
\end{eqnarray}
where $J_{LR}=2t|\lambda|/|\varepsilon|$.  The solution of these equations defines the above expression for the Kondo temperature
and the conductance,
\begin{eqnarray}
  G_{\rm{peak}} =
  \frac{\pi^2e^2}{h}~
  \frac{\lambda^2}{\big|t\big|^2}~
  \frac{W(2\nu+B)}{N(0) T}~
  \frac{1}{\ln^2\big(\frac{d(\nu,B)}{T_K}\big)},
  \label{G-peak}
\end{eqnarray}
where
$$
  d(\nu,B)=
  \Big(
      \big(T^2+4\nu^2\big)
      \big(T^2+(2\nu+B)^2\big)
  \Big)^{1/4}.
$$
Actually the RG equations (\ref{RG}) show that Kondo instability is related to the normal lead while the impact of resonance tunneling through the MBS on $T_K$ is irrelevant. We also notice that the scaling invariance of the exchange part of the conductance  $d\sigma^{ex}/d\ln D=0$  is in agreement  with the second RG equation (\ref{RG}). Thus if $T \leq T_ K$, then, unlike the case $T \gg T_K$, the condition $\Delta \gg T_K$ has to be supplemented by the inequality $\Gamma_S<\Gamma$ in order to justify the small contribution of the continuous  spectrum of the superconductor.

To analyze the strong coupling limit  of the  N-QD-MBS system,  we recall the Anderson hamiltonian for the quantum dot.  In Nambu space the hamiltonian and the current operator acquire the form
\begin{eqnarray}
                H_{TM} &=& \frac{1}{2}[d^{\dagger}\tau_z \hat{V}\gamma_1+t d^{\dagger}\tau_z c(0)]+h.c.\\
                H_d &=& \frac{1}{2}[ \varepsilon d^{\dagger}\tau_z d +U n_{d\uparrow}n_{d\downarrow}]\label{Hd},\,\,\,\hat{I}=\frac{ie}{\hbar}td^{\dagger} c(0)+h.c.\nonumber
              \end{eqnarray}

In the limit $U/t \rightarrow \infty$, employing the slave boson technique, the dot is empty or singly occupied. The elctron creation operator is written as, $d_\sigma^{\dagger}= f_\sigma^{\dagger} b$ where the slave fermion $f_\sigma^{\dagger}$ and the slave boson $b$ mimic the singly occupied and empty dot states. They fulfill the constraint $\sum_{\sigma}d_\sigma^{\dagger}d_\sigma +b^{\dagger}b=1$ that is  encoded by including a Lagrange multiplier $\eta$  the  action $ S $ and replacing the $U$ term in the dot Hamiltonian. At the mean field level the constraint is satisfied only on the average. In the first step, a formula for the average current using the effective action $S_{eff}$ is obtained. This action depends on two c-number parameters: the boson field $b_c$, and the  chemical potential $\eta_c$. In addition,  the action is a function of one quantum source field $\theta_q $ which describes the interaction with the current $\hat{I}$.  Keldysh technique is employed to calculate the  partition function $Z(\theta_q)=\int D(d^{\dagger} d c^{\dagger} c \gamma )\exp[iS]$  that depends on $\theta_q$:
Taking the variation of $ \ln Z$ on $\theta_q\rightarrow 0$ we get expressions for the average current.
  $I = (e/2\hbar) \delta \ln Z/\delta \theta_q$. After replacing the $d^{\dagger},d $ with $(f^{\dagger},f)b_0$
and introducing the constrain into the action, the problem (in MFSB approximation) becomes gaussian. Integrating out the field of the normal lead variables, the quantum dot  slave fermions $f, f^{\dagger}$ and Majorana fermions $\gamma$ we are left with the fermion part of the partition function at
\begin{equation}
  \ln Z_f(\theta_q) = \mbox{tr}\ln \{\frac{-i}{2}(G_f^{-1} - \theta_q \Gamma  b_0^2\tau_z[\tilde{g},\hat{\rho}_x])\}.  \label{Z}\\
  \end{equation}
  where $G_f^{-1}  = G_{f0}^{-1}-b_0^2\hat{\Gamma}$ is the inverse dot total GF.
  We keep  only first order in quantum field $\theta_q$ which is enough to calculate  current.
Here $G_{f}$ and $ \hat{\Gamma}$ are  matrices of the same form in Keldysh space: $ \hat{\Gamma}=\tiny {\left(
\begin{array}{cc}
\Gamma^R & \Gamma^K \\
0 & \Gamma^A \\
\end{array}
\right) }$
with entries as retarded, advanced and Keldysh components. The dot inverse GF $G_{f0}^{-1R}=\sigma_0\otimes(i\partial_t-\tilde{\varepsilon}\tau_z)$  describes the noninteracting  level with energy shift $\varepsilon\rightarrow\tilde{\varepsilon}=\varepsilon+\eta_c$. The  term $\hat{\Gamma}$ of the total quantum dot GF is the principal contribution. We define  $\hat{\Gamma}$ below and derive  an expression for the average current. Differentiate $\ln Z_f$ and performing the trace in Keldysh space we get a compact formula
\begin{equation}\label{I}
    I=\frac{-ie\Gamma b_0^4}{2\hbar}tr[(G_f^R\Gamma^KG_f^A+2G_f^RIm\Gamma^R G_f^A \bar{g}^K)\tau_z].
\end{equation}
where $Im\Gamma^R=(\Gamma^R-\Gamma^A)/2i$. The vertex $\hat{\Gamma}$ acquires a form
\begin{equation}\label{vv}
    \hat{\Gamma}=\Gamma\tilde{g}+\tau_z\hat{V}G_{11}\hat{\rho}_x \hat{V}^{\dagger}\tau_z~.
\end{equation}
Here $G_{11}=\tiny {\left(
         \begin{array}{cc}
              G^K_{11} & G^R_{11}    \\
               G^A_{11} & 0     \label{mkel}\\
               \end{array}
                \right)}$ is the Majorana GF of $\gamma_1$ state.
From Eq.(\ref{vv}) we can obtain all entries
\begin{eqnarray}
  2Im\Gamma^R(\omega) &=& -\Gamma+2|\lambda|^2 ImG_{11}^R(\omega)(1-\hat{\Omega}) \\
  \Gamma^K(\omega) &=& \Gamma\bar{g}^K+2i|\lambda|^2(1-\hat{\Omega}) \tanh\frac{\omega}{2T}ImG_{11}^R \nonumber
\end{eqnarray}
where $\hat{\Omega}=\sigma_y\otimes\tau_z+\sigma_z\otimes\tau_x +\sigma_x\otimes\tau_y$
\begin{figure} [!ht]
\centering
\includegraphics [width=0.4 \textwidth ]{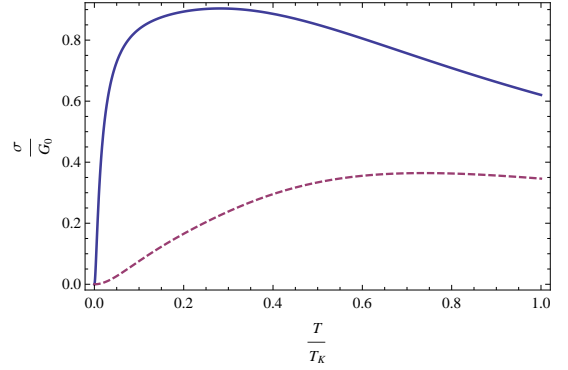}
\caption {\footnotesize {The zero biased conductance as a function of temperature at zero magnetic field in the strong coupling limit ($T<T_K$). The dashed curve corresponds $\nu=0.5T_K$ and the solid line $\nu=0.1 T_K$}.}
\label{Fig4}
\end{figure}

The MFSB approximation is more reliable in equilibrium where $V\rightarrow0$. Therefore, we consider below the temperature dependence of the zero bias conductance for different couplings  between two MBS. In equilibrium the mean field equations for boson field $b_0$ and Lagrangian variable $\eta_q$ can be obtained by minimizing the free energy $F$ on these two parameters.
\begin{equation}\label{F}
    F=-T\sum_{\omega_n}\mbox{tr}\ln[G^{-1}_f(\omega_n)]+\eta_c b_0^2.
\end{equation}
In Eq.(\ref{F}) the last term is the slave boson part  of the free anergy that is the result of constraint. The Matsubara GF of the dot $G_f$  can  be easily be obtained  from $G^R_f(\omega)$. Due to the condition $\Gamma_S<\Gamma$ and $T_K \ll\Delta$  the contribution  of  continuous spectrum above the superconducting energy gap can be neglected. The mean field  equations are
\begin{eqnarray}
  b_0^2 &=& -T \sum_{\omega_n}\mbox{tr}[\tau_zG_f(\omega_n)]+1. \label{eta}\\
  \eta_c &=& -T \sum_{\omega_n}\mbox{tr}[G_f(\omega_n)\hat{\Gamma}(\omega_n)].\label{b0}
\end{eqnarray}
The first equation
(\ref{eta}) fixes the level position $\tilde{\varepsilon}$. In the Kondo
regime the level is singly occupied $\tilde{\varepsilon}=0$ or $\eta_c=|\varepsilon|$. This approximate solution $\eta_c$  of the second equation (\ref{b0}) is used in the first equation to derive nontrivial  solution ($b_0\neq 0$) for the boson filed in terms of Kondo temperature. In the Kondo regime non-logarithmic terms in Eq(\ref{b0}) can be discarded. Direct calculations show that such logarithms appear only for normal metal lead and the all terms related to the MBS can be dropped. Thus we have $b_0^2\Gamma/2=T_K$  where $T_K$  was defined above for the $N$ lead in agreement with the weak coupling analyzis (\ref{RG}). For $T<T_K$ the Eq.(\ref{I}) yields an expression
for the linear conductance
\begin{eqnarray}
  \frac{\sigma}{G_0} &=& g^2_{\Gamma} g^2_ {\lambda}\int_0^{\infty}\frac{dx}{\cosh^2x}\frac{u^2(x)+v^2(x)}{(x^2+g^2_{\Gamma})A(x)}.\label{con}\\
  A(x)&=&x^2+(g_{\Gamma}-g_{\lambda}v(x))^2+g^2_ {\lambda}u^2(x)-2xg_ {\lambda}v(x).\nonumber
\end{eqnarray}
where $G_0=2e^2/h$, and the real functions $u, v$ are defined from $u(x)+iv(x)=x/[(x+i\delta)^2-\tilde{\nu}^2]$. Pertinent dimensionless parameters are: the interaction energy of Majorana fermions $\tilde{\nu}=\nu/T$, the tunneling rates
$g_{\Gamma}=\Gamma b_0^2/4T=T_K/2T$, $ g_ {\lambda}=2|\lambda|^2 b_0^2/T^2$. Figure \ref{Fig4} displays the linear conductance versus temperature  for two values the Majorana fermion coupling energy $\nu$ and $|\lambda|^2=\Gamma T_K/8$. The zero biased conductance as function of temperature displayed in Fig.~\ref{Fig4}, reflects  the similar dependence in the weak coupling limit $T>T_K$ shown in Fig.~\ref{Fig3}. In both cases, if $\nu < T$, smearing starts at lower temperature and correspondingly the peak in the conductance shifts toward lower temperatures.\\

\noindent
\underline{{\it Conclusion}:}
Keldysh technique has been employed to calculate the linear and non-linear conductance in a system consisting of a quantum dot connected to a metal lead on one side and 1D TS hosting Majorana bound states on the other side. The dot was tuned to the Kondo regime. Under certain approximations the whole system is reduced to N-QD-MBS tunneling system in both  the weak ($T \gg T_K$) and strong ($T<T_K$) coupling limits. The conductance has two peak structure if the coupling energy $\nu$   of MBS exceeds the temperature. Under a constant magnetic field,  Ziman splitting occurs on the dot and reduces the heights of the peaks. The magnetic field may result in a more complicated structure of non-linear conductance peaks. Renormalization group analysis is performed in the weak coupling limit while the mean field slave boson approximation is used at $T<T_K$. 
Our analysis shows that in an attempt to probe the features of MBS physics, the role of the Kondo effect  is decisive. 
It is manifested by the occurrence of a strong temperature dependence of the zero bias conductance, and exposes distinct behavior of the non-linear conductance as  compared with 
that of the simpler N-MBS tunnel junction.

\begin{acknowledgments}
We would like to thank  A. Rosch and D. Loss for stimulating discussions.
This research was supported by {\it The Israeli Science Foundation}  grants [No. 1078/07, (AG) and  1703/08 (YA)] .
\end{acknowledgments}

\end{document}